\documentclass{iopconfser}

\usepackage{graphicx}
\usepackage{amsthm,amsmath,amssymb}
\usepackage{subfig}
\usepackage{citesort}
\begin{document}
\nocite{*}

\title{A class of entangled and diffeomorphism-invariant states in loop quantum gravity: Bell-network states}

\author{Bekir Bayta\c{s}$^{1}$,}

\affil{$^1$Department of Mathematics, {\.{I}}zmir Institute of Technology, G{\"{u}}lbah{\c{c}}e, Urla, 35430, {\.{I}}zmir, Turkey}

\email{bekirbaytas@iyte.edu.tr}

\begin{abstract}
Bell-network states constitute a class of diffeomorphism-invariant and entangled states of the
geometry within loop quantum gravity (LQG) that satisfy an area-law for the entanglement entropy in the
limit of large spins. The fluctuations of the geometry for a Bell-network state are entangled, similar to
those in the semiclassical limit as described by quantum field theory in curved spacetimes. We
present a comprehensive analysis of the effective geometry of Bell-network states on a dipole graph. This analysis provides a detailed characterization of the quantum geometry of a class of diffeomorphism-invariant, area-law states representing homogeneous and isotropic configurations in loop quantum gravity, which may be explored as boundary states for the dynamics of the theory. 
\end{abstract}

\section{Introduction}

There are certain observational and theoretical considerations that can support the search for a special class of quantum states within a theory of quantum gravity, where these states serve as a concrete candidate for semiclassical configurations while encoding nontrivial quantum correlations of the geometry. From an observational standpoint, this motivation is particularly relevant in light of the small yet significant anisotropies observed in the cosmic microwave background (CMB)~\cite{Planck:2018vyg}, which may carry imprints of such underlying quantum gravitational structures. These cosmological anisotropies can be interpreted as quantum fluctuations of the geometry of space. In particular, the power spectrum of scalar perturbations scales like the Planck area divided by the Hubble area~\cite{Mukhanov:1981xt}, $\mathcal{P}_s(k) \sim \ell^2_{\mathrm{Pl}} \, H^2_{*}/\epsilon_{*} \sim 10^{-10}\,,$ suggesting that quantum geometry fluctuations, which can be well-described by quantum field theory on curved spaces -as a semiclassical limit of a quantum theory of gravity-, are relevant even at scales much larger than the Planck scale. Phenomenological imprints of Planck-scale dynamics have been explored in cosmological data as a prospective path for detecting signatures of quantum gravitational effects~\cite{Amelino-Camelia:2008aez,Agullo:2013ai,Barrau:2013ula}.  

From a theoretical standpoint, entanglement serves as a valuable diagnostic tool for identifying semiclassical geometries in quantum gravity. The Bianchi–Myers conjecture~\cite{Bianchi:2012ev} posits that the emergence of an area-law scaling for entanglement entropy in a given region $R$, $S_R(| 0 \rangle)= 2 \pi  \frac{\mathrm{Area}(\partial R)}{\ell^2_{\mathrm{Pl}}} + \cdots \,,$ is not a generic feature of arbitrary quantum states, but rather a signature feature of states with semiclassical geometric interpretation. This insight is reinforced by a variety of approaches across quantum gravity and quantum field theory, which consistently point toward area-law behavior~\cite{VanRaamsdonk:2010pw,Jacobson:1995ab,Jacobson:2015hqa,Chirco:2017xjb,Cao:2016mst,Livine:2017fgq}. 

Motivated by these standpoints, we focus on identifying a class of states within the LQG Hilbert space that: (i) exhibit area-law entanglement scaling, (ii) encode average geometries with quantum fluctuations, and (iii) capture nontrivial quantum correlations reminiscent of those in a quantum field. We argue that \textit{Bell-network states}~\cite{Baytas:2018wjd,Bianchi:2018fmq,Baytas:2022mjj,Baytas:2024cux} provide a concrete realization of such LQG states.

The manuscript is organized as follows. Section~\ref{sec:2} begins with a concise review of the kinematical framework of loop quantum gravity restricted to an abstract graph, followed by an overview of the construction of Bell-network states on a generic graph. In Section~\ref{sec:3}, we examine the effective geometric properties of Bell-network states with fixed spins on a dipole graph with four links. Finally, Section~\ref{sec:4} summarizes the main results and conclusion of the work.

\section{Entangled and diffeomorphism-invariant LQG state: Bell-network states}
\label{sec:2}

\subsection{Kinematical setting of loop quantum gravity on a graph}

We consider the kinematical states LQG Hilbert space~\cite{Ashtekar:2004eh,Rovelli:2004tv,Thiemann:2007pyv,Ashtekar:2017yom,Ashtekar:2021kfp} on a single abstract graph $\Gamma = \langle \mathcal{N}, \mathcal{L} \rangle$, where $\mathcal{N}$ and $ \mathcal{L}$ are node and link set of the graph $\Gamma$. The kinematical Hilbert space consists of states on the links, projected onto the subspace invariant under both SU(2) transformations and graph automorphisms~\cite{Baytas:2022mjj}:
\begin{equation}
\otimes_{\ell} \, H_\ell \quad {\underset{P_\Gamma}{\longrightarrow}} \quad \mathcal{H}_{\Gamma} \quad {\underset{P_A}{\longrightarrow}}  \quad K_{\Gamma} \, \cong \, \mathrm{Inv}_{\mathrm{Aut}(\Gamma)}\,\big(L^2[\mathrm{SU}(2)^L/\mathrm{SU}(2)^N]\big) \,.
\end{equation}

%\begin{figure}[htbp]
%\centering
 %\includegraphics[scale=0.35]{spin-network}
%\caption{Illustration of a quantum geometry on an abstract spin-network graph, where each node represents a quantum polyhedron.}
%\label{fig:spin-network}
%\end{figure}

We restrict the set of observables $\mathcal{O}_\Gamma: K_{\Gamma} \to K_{\Gamma}$ defined in a way that they remain unchanged under the action of arbitrary automorphisms $A: \mathcal{N} \to \mathcal{N}$ or SU(2) transformations $g: \mathcal{L} \to \mathrm{SU}(2)$. Concretely, they are introduced by (automorphism) group-averaging the associated operators $\mathcal{O}: H_{\Gamma} \to H_{\Gamma}$ that are defined on a specific presentation of the graph. 
%\begin{equation}
%\mathcal{O}_\Gamma = \frac{1}{| \mathrm{Aut}(\Gamma)|} \sum_{A \in \mathrm{Aut}(\Gamma)} \!\!\! U_A \, \mathcal{O} \, U^{-1}_A \,.
%\end{equation}
For analyzing geometric features of kinematical states $\rho_\Gamma \equiv | \Psi_\Gamma \rangle \langle \Psi_\Gamma | \in K_{\Gamma}$, a guiding practical toolkit is provided as follows~\cite{Baytas:2022mjj}:
\begin{equation}
\rho_\Gamma = U_g \, \rho_\Gamma \, U^{-1}_g = U_A \, \rho_\Gamma \, U^{-1}_A  \,\,\, /  \,\,\,  \mathcal{O}_\Gamma = U_g \, \mathcal{O}_\Gamma \, U^{-1}_g  = U_A \, \mathcal{O}_\Gamma \, U^{-1}_A \,\,\, /  \,\,\, \mathrm{Tr}(\rho_\Gamma \, \mathcal{O}_\Gamma) = \mathrm{Tr}(\rho_\Gamma \, \mathcal{O})\,, \,\,\, \forall \, A, g \,.
\end{equation}
This construction guarantees that both the states and observables are independent of any background structures, including specific graph labels. We assert that Bell-network states are defined within this framework, satisfying both SU(2) gauge invariance and invariance under graph automorphisms.

\subsection{Bell-network states on a generic graph}

In LQG, spin-network basis states $|\Gamma, i_n, j_\ell \rangle$~\cite{Rovelli:1994ge} and coherent states $|\Gamma,j_{\ell},\Phi_n(\texttt{n}_{n,\ell})\rangle$~\cite{Livine:2007vk,Freidel:2010aq} are product states of intertwiners $|i_n \rangle $ and coherent intertwiners $|\Phi_{n}(\texttt{n}_{n,\ell})\rangle$ with uncorrelated quantum fluctuations, respectively. Each intertwiner state $|i_n \rangle $ represents a quantum polyhedron. We look for a corner of the Hilbert space where quantum correlations are present. To construct such states, we introduce a local state $|\mathcal{B},\lambda_{\ell}\rangle \in H_{\ell}$ for each link $\ell \in \mathcal{L}$ with $\lambda_\ell \in \mathbb{C}$, which is inspired by squeezed vacuum techniques~\cite{Bianchi:2016hmk}:
\begin{align}
|\mathcal{B},\lambda_{\ell}\rangle &= (1 - |\lambda_{\ell}|^2) \, \mathrm{exp}\big(\lambda_{\ell} \, \epsilon_{B B'} \, a^{B \dagger}_{s(\ell)} \, a^{B' \dagger}_{t(\ell)}\big) \, |0\rangle_{s(\ell)} |0\rangle_{t(\ell)} \\
& = (1-|\lambda_{\ell}|^2) \sum_{j_{\ell}} \, \lambda_{\ell}^{2j_{\ell}} \!\!\! \sum_{m_{\ell}=-j_{\ell}}^{+j_{\ell}} (-1)^{j_{\ell}-m_{\ell}} |j_{\ell},m_{\ell}\rangle_{t(\ell)} \, |j_{\ell},-m_{\ell}\rangle_{s(\ell)} \, ,
\end{align}
which is given as a superposition of the generalized Bell state, a maximally entangled state of spin $j_\ell$. From a information-theoretic point of view, the underlying principle of producing these local states is really just to maximize the mutual information between source $s(\ell)$ and target $t(\ell)$ nodes at each link $\ell$~\cite{Baytas:2018wjd}.

By taking the tensor product of such link-states $|\mathcal{B},\lambda_{\ell}\rangle$ over the graph $\Gamma$ and projecting to the SU(2)-invariant subspace $\mathcal{H}_\Gamma$, $|\Gamma,\mathcal{B},\lambda_{\ell}\rangle \equiv P_{\Gamma}\otimes_{\ell} |\mathcal{B},\lambda_{\ell}\rangle$, we obtain a state in $\mathcal{H}_\Gamma$, which we call a Bell-network state $|\Gamma,\mathcal{B},\lambda_{\ell}\rangle$ on $\Gamma$. When $\lambda_\ell = \lambda, \, \forall \ell$, the resulting state is invariant under graph automorphisms: $U_A |\Gamma,\mathcal{B},\lambda\rangle = |\Gamma,\mathcal{B},\lambda \rangle \,, \, \forall A \in \mathrm{Aut}(\Gamma)$~\cite{Baytas:2022mjj}. 

The projection $P_\Gamma$ can be implemented using the resolution of the identity in the spin-network basis, leading to a formula for the Bell-network state $ |\Gamma,\mathcal{B},\lambda \rangle $ as an expansion over spin configurations~\cite{Baytas:2018wjd,Baytas:2022mjj}:
\begin{equation}
|\Gamma,\mathcal{B},\lambda \rangle = \frac{1}{\sqrt{N}} \sum_{j_{\ell}}\bigg(\prod_{\ell} \sqrt{2 j_{\ell}+1} \, \lambda^{2 j_{\ell}} \bigg) \sum_{i_n} \, \overline{\mathcal{A}_{\Gamma}(j_{\ell},i_n)} \, |\Gamma, i_n, j_\ell \rangle \,, \,\,\, \mathcal{A}_{\Gamma}(j_{\ell},i_n) = \sum_{\{m\}} \prod_n [i_n]^{m_1 \cdots m_{F_n}} \, ,
\end{equation}
where the amplitude $\mathcal{A}_{\Gamma}(j_{\ell},i_n)$ is the $\mathrm{SU}(2)$-symbol of the graph $\Gamma$ and the contraction of the intertwining tensors is according to the combinatorics of the graph. The entanglement ensures that adjacent polyhedra have compatible normals, enforcing a form of geometric continuity not found in product states. It is that the presence of the strong correlations of the fluctuations of the geometry tame the induced discontinuities of geometries from generic states in LQG, in particular restricting the fluctuations of the geometry to a subspace of vector geometries~\cite{Barrett:2009gg,Dona:2017dvf}, in which the normals of neighboring polyhedra are glued back-to-back. This gluing behavior is in fact important— it is what allows the state to encode semiclassical geometry, by representing a superposition over glued polyhedra. 

\section{Effective geometry of Bell-network states on a dipole graph}
\label{sec:3}

\subsection{States and observables on a dipole graph}

Let us provide a detailed characterization of the effective geometries of Bell-network states on a dipole graph $\Gamma_{2,L}$, which is minimal but nontrivial configuration. States that are defined on $\Gamma_{2,L}$ are used for representing homogeneous and isotropic configurations. Its automorphism group consists of permutations of the two nodes and $L$ links: $\mathrm{Aut}(\Gamma_{2,L}) = \mathcal{S}_{2} \times \mathcal{S}_L$ and $|\mathrm{Aut}(\Gamma_{2,L})| = 2! L!$. The invariant observables of interest are obtained by using group-averaged local geometric quantities: volume at a node, area on a link and dihedral angles at a wedge~\cite{Baytas:2024cux}:
 \begin{equation}
 \mathcal{V} = \frac{1}{2} \sum_{n} \mathcal{V}_n\,, \quad \mathcal{A} = \frac{1}{2 L} \sum_{\ell} \mathcal{A}_{\ell}\,, \quad 
\cos \Theta = \frac{1}{2 L(L-1)} \sum_{w} \cos \theta_w \,. 
 \end{equation}
These observables behave like one-body operators familiar from many-body physics~\cite{Martin}, but now invariant under graph symmetries. So, this setup allows us to extract effective geometric data from Bell-network states. What is interesting is that even this simple graph is enough to witness nontrivial quantum gravitational features.

\subsection{Bell-network states on a dipole graph}

On the dipole graph with four links, a Bell-network state at fixed spins displays perfect correlations
between the two nodes~\cite{Baytas:2018wjd,Baytas:2024cux}:
\begin{equation}
|\Gamma_{2,4},\mathcal{B},j_\ell \rangle := P_A \, P_{j_\ell} \, |\Gamma_{2,4},\mathcal{B}, \lambda \rangle = \frac{1}{4!} \,\sum_\sigma \frac{1}{\sqrt{\mathrm{dim} H(j_\ell)}} \sum_{k=1}^{\mathrm{dim} H(j_\ell)} | j_{\sigma(\ell)}, i_k, (\zeta i)_k \rangle \,,
\end{equation}
where $\mathrm{dim} H(j_\ell)$ is the dimension of the intertwiner space and the anti-linear map $\zeta$ relates the intertwiners at source and target nodes — so observables at both nodes yield identical outcomes. This mirrors EPR-like correlations in geometry, allowing us to speak of entangled polyhedra
with matching intrinsic geometry.

\subsection{Effective geometry as spherical tetrahedron}

\begin{figure}[hbtp]
    \centering
    \subfloat[\centering $D_{\mathcal{B}}(j_\ell)|_{j_1 = j_2 = j_3 = j, j_4= 1/2}$]{{\includegraphics[scale=0.2]{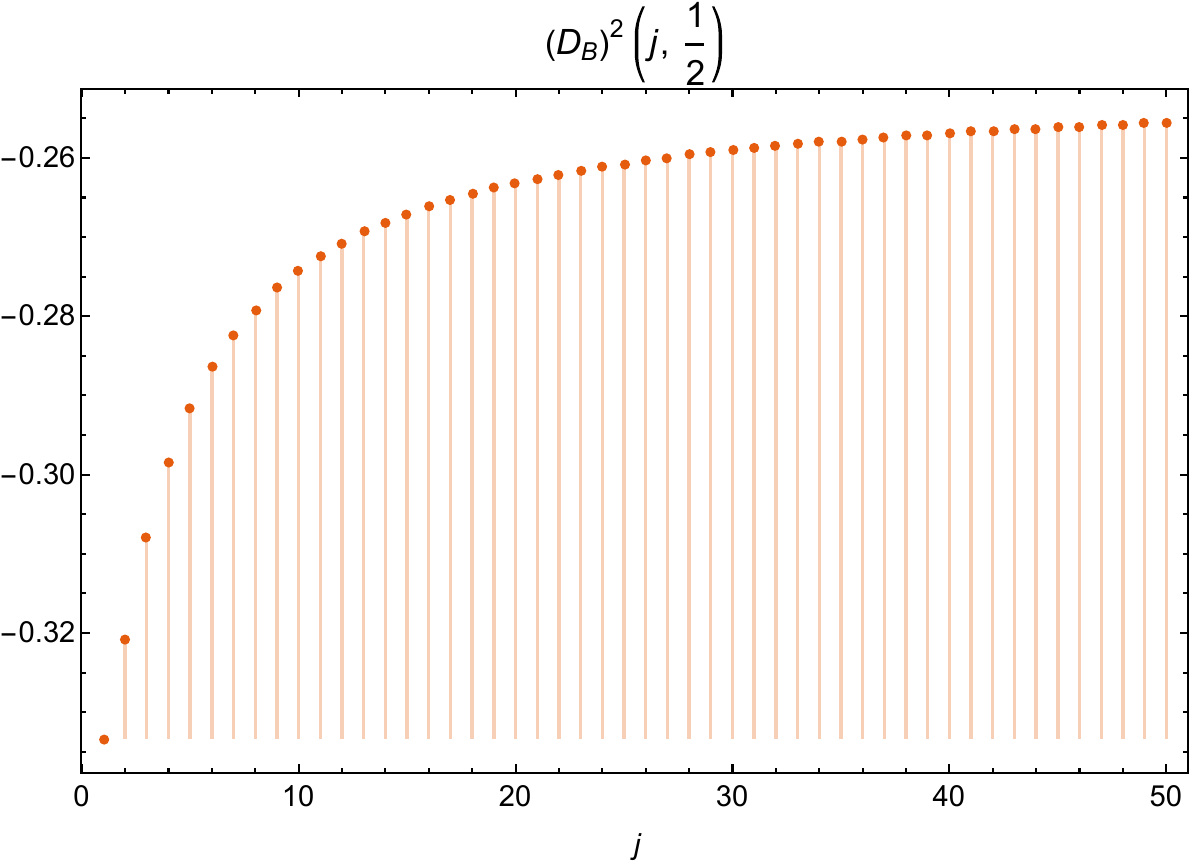} }}%
    \qquad
    \subfloat[\centering $\langle \mathcal{V} \rangle_{\mathcal{B},j}$]{{\includegraphics[scale=0.2]{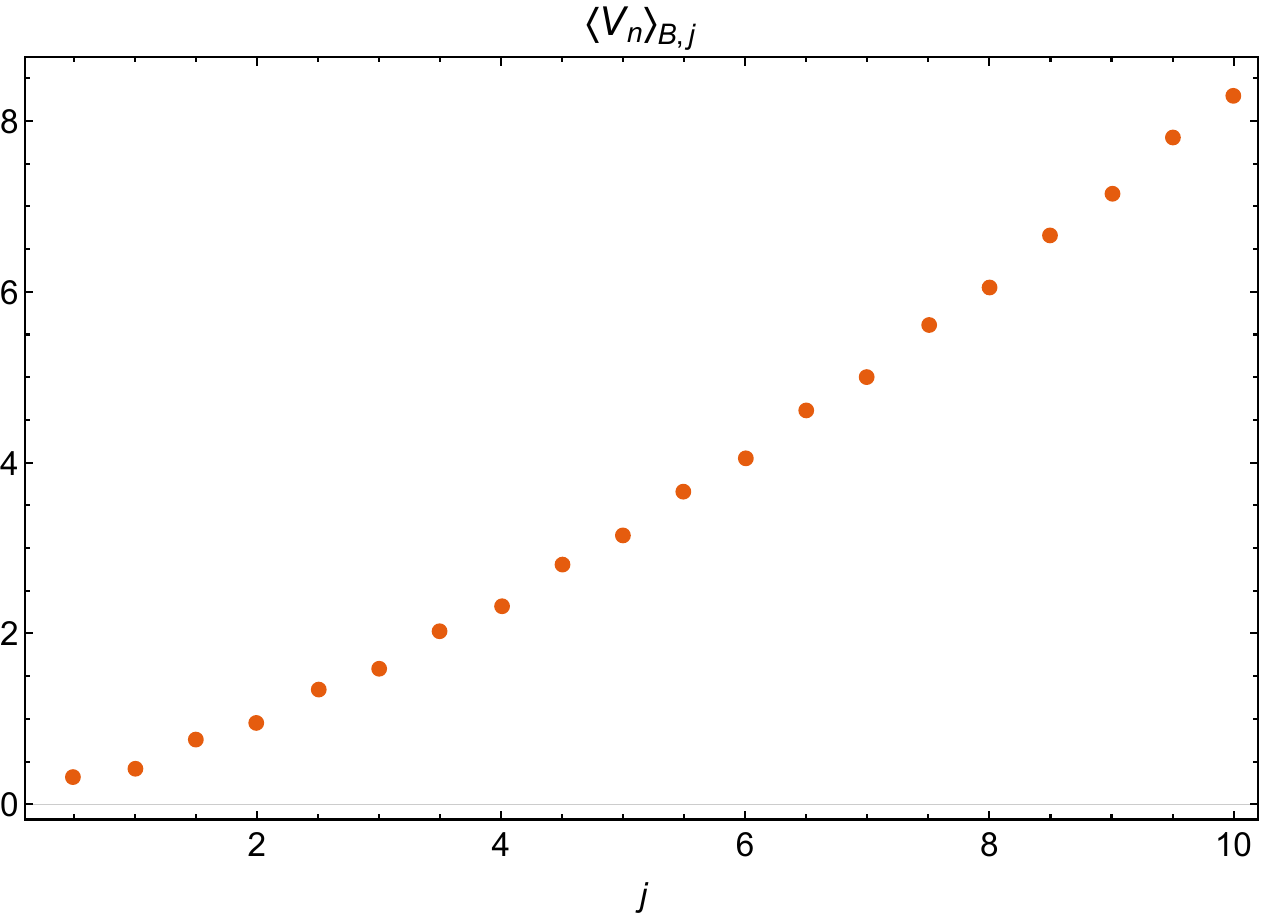} }}%.
      \qquad
      \subfloat[\centering $\Delta(\cos \Theta)_{\mathcal{B},j_\ell=j}$]{{\includegraphics[scale=0.2]{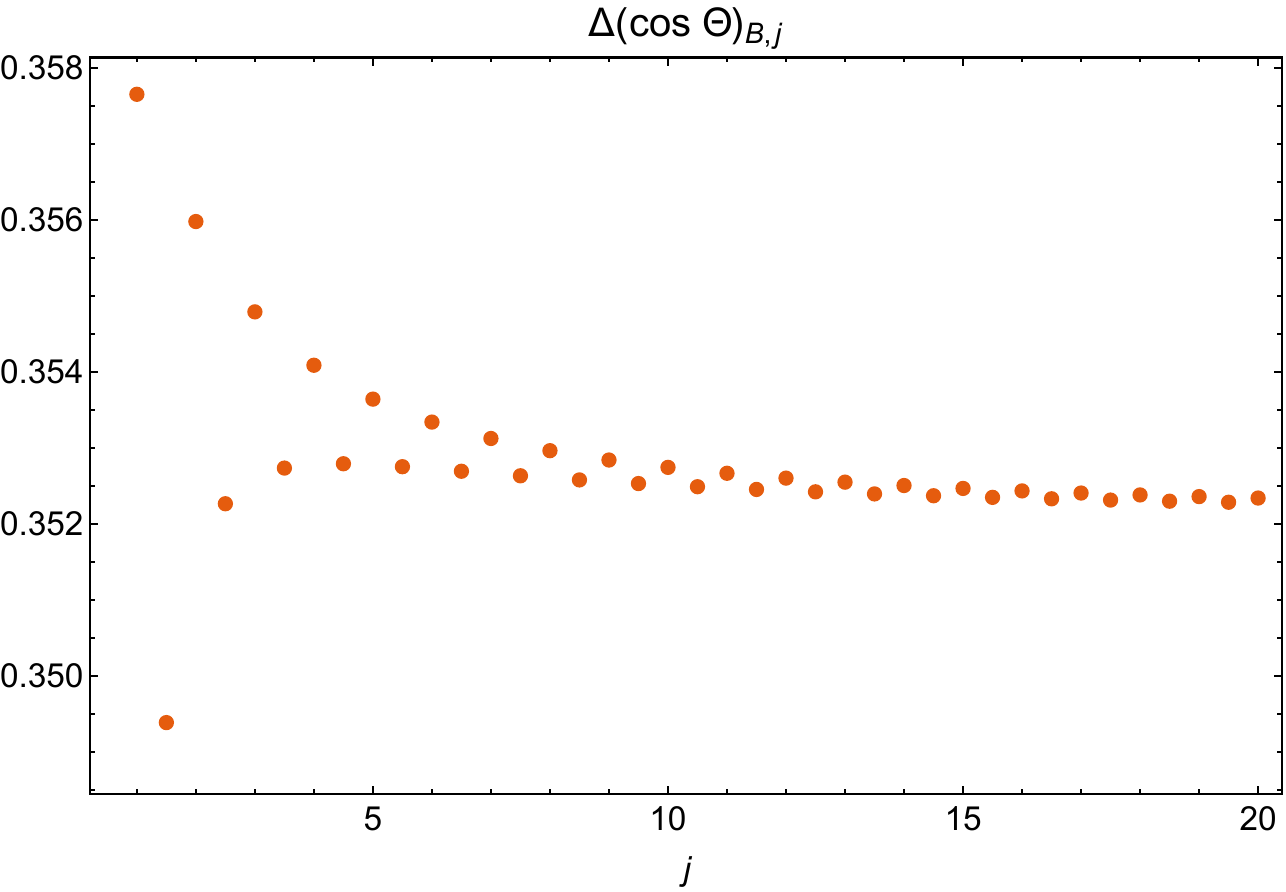} }}%.
    \caption{Expectation values and dispersion of $\{\cos \Theta, \mathcal{V}\}$ relative to $|\Gamma_{2,4},\mathcal{B},j_\ell \rangle$.}%
    \label{fig:computations}%
\end{figure}

We analyze the expectation values of volume, area and dihedral angles relative Bell-network state at fixed spins $j_\ell$: $ \{\langle \mathcal{V} \rangle_{\mathcal{B},j_\ell} \,, \, \langle \mathcal{A} \rangle_{\mathcal{B},j_\ell} \,, \, \langle \cos \Theta \rangle_{\mathcal{B},j_\ell}\}$~\cite{Baytas:2024cux} (See Fig.~\ref{fig:computations}). $D_{\mathcal{B}}(j_\ell) := \langle \cos \Theta \rangle_{\mathcal{B}, j_\ell}$ describes the average of the cosine of the dihedral angle in any pair of faces at one of the nodes. The sign of the constant curvature can be determined by computing the determinant of the Gram matrix associated with the regular tetrahedron, $
\mathrm{det} \, G_{j_\ell} := (1 - D_{\mathcal{B}}(j_\ell))^3 \, (1 + 3  D_{\mathcal{B}}(j_\ell))$. When all spins are equal $j_\ell = j_0$, the average geometry corresponds to a regular flat tetrahedron: $D_{\mathcal{B}}(j_0) = -1/3 \,, \,\,\, \mathrm{det} \, G_{j_0}= 0$. When spins differ, the geometry becomes that of a regular spherical tetrahedron, $D_{\mathcal{B}}(j_\ell) > -1/3 \,, \,\,\, \mathrm{det} \, G_{j_\ell} > 0$, indicating spatial curvature~\cite{Baytas:2024cux}. Even in the large-spin limit, dihedral angle fluctuations $\Delta(\cos \Theta)_{\mathcal{B},j_\ell=j}$ remain finite — highlighting persistent quantum fluctuations (See Fig.~\ref{fig:computations}).
\section{Summary} 
\label{sec:4}
Bell-network states are entangled, automorphism-invariant states and live in the LQG Hilbert space. They enforce gluing conditions through entanglement and support an area law for entropy. They solely depend on the combinatorial structure of the graph, without any additional structure or classical background data. They can model curved geometries, and even serve as boundary states in spinfoam cosmology~\cite{Gozzini:2019nbo,Frisoni:2023lvb}. These features make them compelling semiclassical candidates. 

\section*{Acknowledgments}

B.B. acknowledges Eugenio Bianchi and Nelson Yokomizo for useful feedbacks.

\bibliography{iopart-num}

\end{document}